\documentclass[journal]{IEEEtran}

\ifCLASSINFOpdf
\else
\fi
\usepackage{amsmath}
\usepackage{graphicx}
\usepackage{graphicx}
\usepackage{subcaption}
\usepackage{xcolor}
\usepackage[bottom]{footmisc}
\usepackage{epstopdf}
\usepackage{algorithm}
\usepackage[noend]{algpseudocode}

\makeatletter
\def\BState{\State\hskip-\ALG@thistlm}
\makeatother

\begin{document}
\raggedbottom
%
\title{A framework for mitigating zero-day attacks in IoT}
%
%
%

\author{\IEEEauthorblockN{Vishal Sharma\IEEEauthorrefmark{1}, Jiyoon Kim\IEEEauthorrefmark{2}, Soonhyun Kwon\IEEEauthorrefmark{3}, Ilsun You\IEEEauthorrefmark{4}, Kyungroul Lee\IEEEauthorrefmark{5}, Kangbin Yim\IEEEauthorrefmark{6}} \\
    \IEEEauthorblockA{\IEEEauthorrefmark{1}\IEEEauthorrefmark{2}\IEEEauthorrefmark{3}\IEEEauthorrefmark{4}\IEEEauthorrefmark{5}\IEEEauthorrefmark{6}Department of Information Security Engineering, Soonchunhyang University, \\Asan-si 31538, The Republic of Korea \\
        Email: vishal\_sharma2012@hotmail.com, \{74jykim, tnsgus08, ilsunu\}@gmail.com},\\ \{carpedm, yim\}@sch.ac.kr  \\
}

\maketitle

\begin{abstract}
Internet of Things (IoT) aims at providing connectivity between every computing entity. However, this facilitation is also leading to more cyber threats which may exploit the presence of a vulnerability of a period of time. One such vulnerability is the zero-day threat that may lead to zero-day attacks which are detrimental to an enterprise as well as the network security. In this article, a study is presented on the zero-day threats for IoT networks and a context graph based framework is presented to provide a strategy for mitigating these attacks. The proposed approach uses a distributed diagnosis system for classifying the context at the central service provider as well as at the local user site. Once a potential zero-day attack is identified, a critical data sharing protocol is used to transmit alert messages and reestablish the trust between the network entities and the IoT devices. The results show that the distributed approach is capable of mitigating the zero-day threats efficiently with 33\% and 21\% improvements in terms of cost of operation and communication overheads, respectively, in comparison with the centralized diagnosis system.
\end{abstract}


%
\IEEEpeerreviewmaketitle

\section{Introduction}
The communication networks are observing a tremendous increase in the number of devices which are predicted to go beyond 40\% (of that were active in 2012) by 2020~\cite{zhang2014sybil}. All these devices have been arranged under a common term of ``Internet of Things" (IoT). IoT allows integration of the vast variety of communication devices irrespective of their operational technology, which is also a challenging issue as a common firmware is required for all the devices. A common firmware makes it easier to control and manage various IoT devices without many overheads. Common software platforms allow easy configurations as well as easy diagnosis of faulty operations. However, a common firmware also subjects the IoT components to various types of threats which can infiltrate the operational defense of these devices~\cite{sharma2017saca}. Some of the key features required by IoT networks are remote diagnosis and management, data analytic, software upgrades, information passing and processing, and user mobility identification~\cite{sadeghi2015security}. All these form a type of application which allows access to the entire network once a particular feature is exploited.

Since there is no formal definition of IoT, same attacks which are applicable to any computing entity hold true in their case. Also, reduction in the human interventions and use of more automated systems in the IoT networks make it extremely important to secure the entire network as it may reveal critical information~\cite{kotenko2012attack}. Apart from these, IoT networks are also considered as an integral part of civilian and military expeditions focusing surveillance, navigation, localization, equipment control, and currency transfers, etc. Recent trends have focused on using RFID tags as embedded sources for IoT devices that do not connect to the network directly. Although, such strategy holds safe for the majority of application scenarios, but manipulation with RFID tags can easily make these vulnerable similar to a normal computing entity~\cite{covington2013threat}. Thus, security of IoT devices irrespective of the mode and type of connectivity is of utmost importance and has been an area of concern for a majority of the security researchers across the globe.

Considering a common platform for IoT devices, most of the business enterprises and vendors focus on making version-based IoT firmware that can be easily upgraded and controlled. Such scenarios are possible by using a software-assisted networking. However, a software-assisted networking suffers from a major issue of zero-day vulnerabilities. Considering the level of deployment and configuration of networks, zero-day vulnerabilities are extremely dangerous for IoT networks. Exploitation of a zero-day vulnerability can lead to a zero-day attack~\cite{kaur2014survey}. Control over a single unit of IoT software may expose the entire architecture.
\begin{figure}[!ht]
  \centering
  \includegraphics[width=230px]{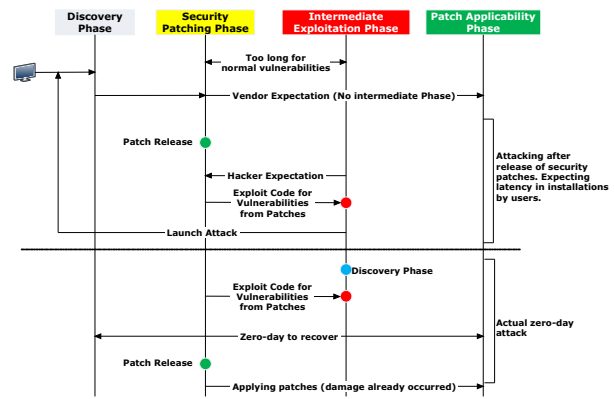}
  \caption{An illustration of the window of vulnerability for zero-day attacks.}\label{fig1}
\end{figure}

\section{Background: zero-day attacks}
The name ``Zero-day" is coined considering the negligible time available in mitigating these threats. The number of days for which an anomaly has been known directly affects the countermeasures and also the probability of remaining affected. It also has to do a lot with those software users who do not update security patches regularly. Once a vulnerability is publicized, it is mandatory for the particular application users to immediately switch to the stable releases. However, failure in doing so leads to various consequences in the form of cyber attacks~\cite{portokalidis2006argos}.

The effect of a zero-day vulnerability also depends on the mode of detection. If a vulnerability is identified by white hat hackers, it allows keeping it low profile until the security patches are not available; whereas identification of such vulnerabilities by a notorious group (black hat hackers) may subject the entire enterprise to failure~\cite{kaur2014survey}. The vulnerability cycle for a zero-day attack may vary from scenario to scenario. In some cases, after identification of a bug, the hackers operate covertly leading to the full zero-day attack, while in some cases, the hackers may come forward (overt) and make threat public~\cite{palani2016invisible}~\cite{wanswett2015threat}. Thus, it can be analyzed that a zero-day attack is not only because of the covert behavior of a hacker but also because of the delays in updating security patches once these are available in the public domain. This is often explained in the terms of window of vulnerability. The window of vulnerability is the time gap in which the number of vulnerable systems remaining is negligible. It is evaluated as a software timeline considering the discovery phase, security patching, intermediate exploitation phase and patch applicability phase, as shown in Fig.~\ref{fig1}~\cite{palani2016invisible}~\cite{bilge2012before}.

\begin{figure}[!ht]
  \centering
  \includegraphics[width=230px]{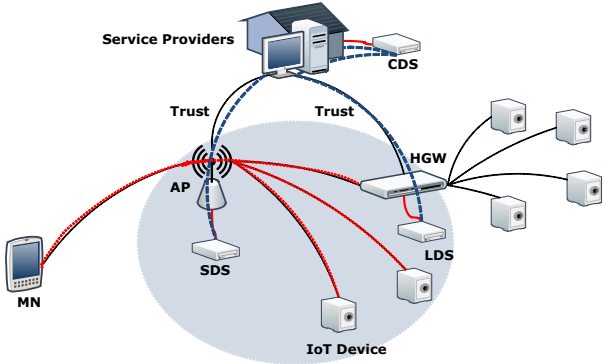}
  \caption{An illustration of DDS-assisted IoT network.}\label{fig3}
\end{figure}

\begin{figure}[!ht]
  \centering
  \includegraphics[width=230px]{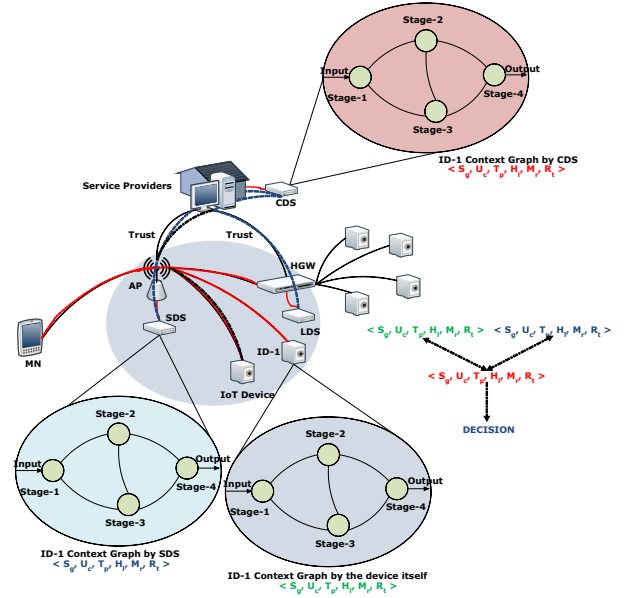}
  \caption{An illustration of strategic context graph formation for an IoT device between the SDS and CDS. The decision on matching context is performed at CDS. The counter updates and firmware version decisions are also evaluated at the SDS and the CDS.}\label{fig4}
\end{figure}

\section{Proposed Approach}
The network comprises various IoT devices and gadgets that operate either individually or collectively via a common gateway. The communication can be directly between the Mobile Node (MN) and the IoT device or indirectly between the MN and the IoT device via a gateway. The service providers are responsible for maintaining trust between the IoT and the MN. Currently, the proposed model emphasizes on a particular scenario in which an IoT device receives security updates that may lead to zero-day attacks; or when an attack is already launched and security updates confirm the attacks. The proposed approach uses strategic context graphs to ensure the safety of IoT devices against the zero-day attacks. The context graphs are implemented using Distributed Diagnosis System (DDS). The DDS are divided into three parts (shown in Fig.~\ref{fig3}), namely,
\begin{itemize}
  \item Central Diagnosis System (CDS): CDS is installed by the service providers on the central node of the network which is responsible for generating trust as well as the updates for the entire network. CDS is responsible for managing the Access Points (APs) control, and the operations of gateways for maintaining security in the case of high possibilities of threats.
  \item Local Diagnosis System (LDS): LDS is operated as a dedicated device over the gateways. Usually, these are installed with the Home Gateways (HGW). LDS interacts with the CDS and shares its context graphs with it to ensure that all the security procedures are followed by the corresponding IoT device.
  \item Semi Diagnosis System (SDS): SDS is responsible for directly managing the APs trust with the CDS. It shares the context of IoT devices which directly interacts with an MN without relying on the local gateway.
\end{itemize}
\subsection{Strategic Context}
The types of devices operable in a network are considered to have valid pre-registered signatures along with a counter value. The counter value manages the count for the number of times the firmware of an IoT device is validated or encountered. The context for each IoT device is managed by its diagnosis system and periodically stored in logs and shared with the CDS. The context outline used in the proposed model is as follows:
\begin{itemize}
  \item Device signatures ($S_g$): This is the unique identity for each device. The signature is the embedded information about the IoT device which is stored at the CDS once it gets activated in the network.
  \item Update Counter ($U_c$): This is the firmware update counter which is randomly selected at the beginning of network registrations. These are updated using random integer values which are finalized by the CDS and change periodically without affecting the performance.
  \item Traffic Type ($T_p$): This defines the context for the type of traffic to be generated for and by an IoT device. This helps the diagnosis system to analyze the content over a particular channel for its correctness.
  \item Header Length ($H_l$): It defines the bit length of the header field used by the diagnosis system. It contains all the necessary context metadata which is to be shared between the LDS, SDS, and CDS.
  \item Memory Range ($M_r$): It denotes the maximum and minimum size of the packets generated by the IoT device. This helps to simply analyze if the size of the initial code is affected or not. Usually, these are not mishandled by the attackers, but still, in some cases, this is very useful to identify if the binaries of the firmware are altered or not.
  \item Route ($R_t$): This field is used to check whether an IoT device is operable in LDS, SDS, or CDS region. This also allows tracking the actual route for managing the context between the network entities.
\end{itemize}
\subsection{Context Graphs and Strategic Attack Detection}
The context graphs are used to generate the strategies which help in taking a decision regarding the presence of a threat amongst the IoT devices. The number of vertices in the context graphs is equal to the number of processing procedures an IoT device follows before generating an output and demanding an input. The context explained above forms the edges of the graph. After the time instance decided in the configuration of the network, the LDS and SDS evaluate these graphs for every corresponding IoT device ad share it with the CDS which also forms its own context graph for every IoT device. Along with the context graphs, the CDS also forms the context graphs for the subordinate network which includes the layers of APs, and gateways.

In order to take a strategic decision on the management of IoT devices against the zero-day attacks, the CDS follows a principle of modeling the counter and the random integer value used to manage the counter by the LDS, SDS and the device itself. Then, it performs mutual exclusion rule to trace the presence of a zero-day threat in the IoT network. The failure in the matching of the context stored and the context received from all the subordinates as well as the IoT device indicates the presence of a zero-day attack. The operational view of the proposed approach is illustrated in the Fig.~\ref{fig4}.

It is to be noted that the strategic context graphs are applicable in the network only in the deployment phase, but not in the development phase. Thus, the proposed strategy can come handy only when a vulnerability is identified by the development team at lateral stages as well as during the release of security updates as it helps in tracking the contextual behavior of every IoT device. Once a possibility of attack is found, the proposed approach utilizes the critical data sharing protocol that helps in eliminating a particular IoT device before it exploits the entire network.
\begin{figure}[!ht]
  \centering
  \includegraphics[width=230px]{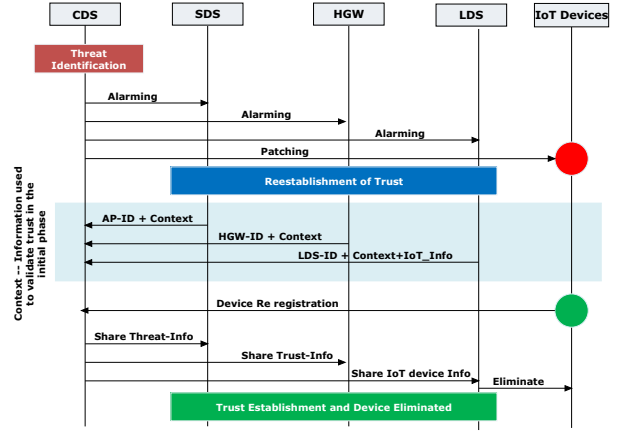}
  \caption{An illustration of critical context/data sharing protocol used after the identification of potential zero-day threat or attack in the IoT network.}\label{fig5}
\end{figure}

\subsection{Critical Data Sharing Protocol}
The proposed approach uses a critical context/data sharing protocol in the scenarios with a potential zero-day threat. The protocol, shown in Fig.~\ref{fig5}, illustrates the procedures opted by the CDS once a threat is identified amongst the IoT devices leading to a zero-day exploitation.

Once a threat policy is violated, the CDS sends alarming messages to its connected components that are its subordinates in the network. The alarming messages are followed by the patch for fixing the affected IoT device. This is followed by the reestablishment of the trust between all the connected components with the CDS. Once an alarming request is received, each subordinate's diagnosis system shares context information to revalidate the trust. By the time, these steps are performed, the affected device updates its security mechanisms, and registers itself again with the CDS leading to the elimination of the threat without eliminating the device. On the contrary, CDS shares threat information with the SDS, trust information with the HGW, device information with the LDS, and finally, leads it to eliminate the incorrect device. This allows mitigating zero-day threats in IoT networks.

\section{Performance Evaluation}
\begin{figure}
\centering
   \begin{subfigure}[b]{0.20\textwidth}
   \centering
  \includegraphics[width=115px]{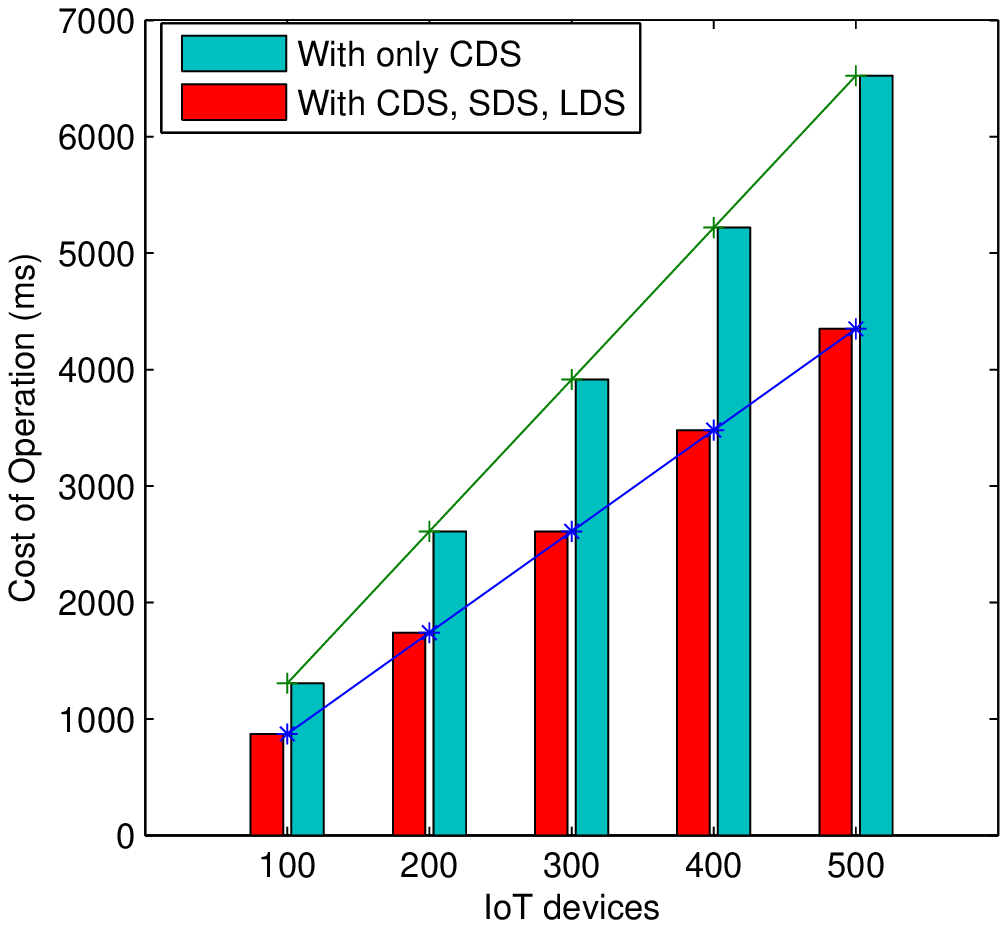}
   \caption{Cost of Operation vs. IoT devices.}
   \label{g1}
\end{subfigure}
\begin{subfigure}[b]{0.20\textwidth}
\centering
  \includegraphics[width=115px]{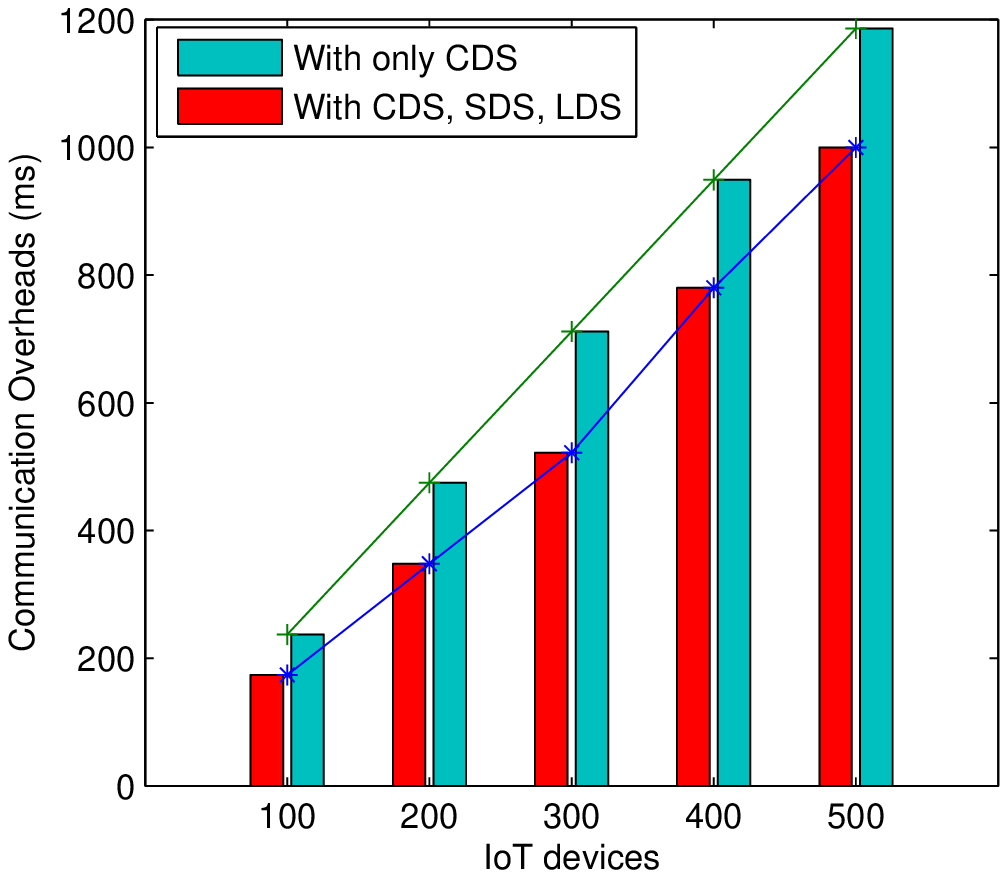}
   \caption{Communication Overheads vs. IoT devices.}
   \label{g2}
\end{subfigure}
\caption{Simulation Results.}
\end{figure}

The proposed approach is evaluated by deploying 500 sensors in two modes, namely, with CDS only and with CDS, LDS, and SDS. The proposed approach is evaluated to analyze the effect of DDS on the performance of the proposed framework. The model defined in Ref.~\cite{sharma2017saca} with similar attacker scenario (20\% nodes as the attacker) is used to evaluate the formation in the proposed approach for cost of operation and communication overheads. The cost of operation is calculated as the time required by the diagnosis system to arrive at the decision of zero-day possibility. It includes the communication time including the context sharing procedures as well as the formation of the context graphs at the interacting entities of the network. Results in Fig.~\ref{g1} show that the DDS is capable of performing better in distributed mode rather than only CDS scenario, and covers 33\% less cost of operation. With critical protocol coming into play after the identification of a zero-day threat, the proposed approach utilizes series of steps to generate alert messages and reestablish the trust between the connected devices and gateways. The DDS causes 21\% lesser overheads in comparison with the scenario with a single diagnosis system, as shown in Fig.~\ref{g2}.
\section{Conclusion}
In this article, a study was presented on zero-day threats for IoT networks. A context graph based framework was presented to provide a strategy for deciding on the zero-day attacks. The proposed approach used a distributed diagnosis system for classifying the context at the central service provider as well as at the local user site. Also, once a zero-day attack was potentially identified, a critical data sharing protocol was used to transmit alert messages and reestablish the trust between the network entities and the IoT devices.

This is a progressive paper and the details on the full-fledged implementation along with critical evaluations will be presented in future reports. 
\bibliographystyle{ieeetr}
\bibliography{final_ref}

\begin{thebibliography}{10}

\bibitem{zhang2014sybil}
K.~Zhang, X.~Liang, R.~Lu, and X.~Shen, ``Sybil attacks and their defenses in
  the internet of things,'' {\em IEEE Internet of Things Journal}, vol.~1,
  no.~5, pp.~372--383, 2014.

\bibitem{sharma2017saca}
V.~Sharma, J.~D. Lim, J.~N. Kim, and I.~You, ``Saca: Self-aware communication
  architecture for iot using mobile fog servers,'' {\em Mobile Information
  Systems}, vol.~2017, 2017.

\bibitem{sadeghi2015security}
A.-R. Sadeghi, C.~Wachsmann, and M.~Waidner, ``Security and privacy challenges
  in industrial internet of things,'' in {\em Design Automation Conference
  (DAC), 2015 52nd ACM/EDAC/IEEE}, pp.~1--6, IEEE, 2015.

\bibitem{kotenko2012attack}
I.~Kotenko and A.~Chechulin, ``Attack modeling and security evaluation in siem
  systems,'' {\em International Transactions on Systems Science and
  Applications}, vol.~8, pp.~129--147, 2012.

\bibitem{covington2013threat}
M.~J. Covington and R.~Carskadden, ``Threat implications of the internet of
  things,'' in {\em Cyber Conflict (CyCon), 2013 5th International Conference
  on}, pp.~1--12, IEEE, 2013.

\bibitem{kaur2014survey}
R.~Kaur and M.~Singh, ``A survey on zero-day polymorphic worm detection
  techniques,'' {\em IEEE Communications Surveys \& Tutorials}, vol.~16, no.~3,
  pp.~1520--1549, 2014.

\bibitem{portokalidis2006argos}
G.~Portokalidis, A.~Slowinska, and H.~Bos, ``Argos: an emulator for
  fingerprinting zero-day attacks for advertised honeypots with automatic
  signature generation,'' in {\em ACM SIGOPS Operating Systems Review},
  vol.~40, pp.~15--27, ACM, 2006.

\bibitem{palani2016invisible}
K.~Palani, E.~Holt, and S.~Smith, ``Invisible and forgotten: Zero-day blooms in
  the iot,'' in {\em Pervasive Computing and Communication Workshops (PerCom
  Workshops), 2016 IEEE International Conference on}, pp.~1--6, IEEE, 2016.

\bibitem{wanswett2015threat}
B.~Wanswett and H.~K. Kalita, ``The threat of obfuscated zero day polymorphic
  malwares: An analysis,'' in {\em Computational Intelligence and Communication
  Networks (CICN), 2015 International Conference on}, pp.~1188--1193, IEEE,
  2015.

\bibitem{bilge2012before}
L.~Bilge and T.~Dumitras, ``Before we knew it: an empirical study of zero-day
  attacks in the real world,'' in {\em Proceedings of the 2012 ACM conference
  on Computer and communications security}, pp.~833--844, ACM, 2012.

\end{thebibliography}








\end{document}